\documentstyle[epsfig,color,12pt]{article}

\include{epsf}

\newcount\timecount
\newcount\hours \newcount\minutes  \newcount\temp \newcount\pmhours
\hours = \time \divide\hours by 60 \temp = \hours \multiply\temp
by 60 \minutes = \time \advance\minutes by -\temp
\def\hour{\the\hours}
\def\minute{\ifnum\minutes<10 0\the\minutes
            \else\the\minutes\fi}
\def\clock{
\ifnum\hours=0 12:\minute\ AM \else\ifnum\hours<12 \hour:\minute\
AM
      \else\ifnum\hours=12 12:\minute\ PM
            \else\ifnum\hours>12
                 \pmhours=\hours
                 \advance\pmhours by -12
                 \the\pmhours:\minute\ PM
                 \fi
            \fi
      \fi
\fi }

\def\monthname{\relax\ifcase\month 0/\or January\or February\or
   March\or April\or May\or June\or July\or August\or September\or
   October\or November\or December\else\number\month/\fi}

\def\bold#1{\setbox0=\hbox{$#1$}%
     \kern-.025em\copy0\kern-\wd0
     \kern.05em\copy0\kern-\wd0
     \kern-.025em\raise.0433em\box0 }

\def\beq{\begin{equation}}
\def\eeq{\end{equation}}

\newlength{\capindent}
\newlength{\capwidth}
\newlength{\figwidth}
\newcommand{\EEGG}{\rm e^+ e^-\rightarrow \gamma\gamma(\gamma)}

\begin{document}

\begin{flushright}
arXive:0907.0629 [hep-ph] \\
CERN-PH-TH/2009-105 \\
June 2009
\end{flushright}

\centerline{\Large\bf Minimal Length Scale in Annihilation}

\vskip 0.2in

\centerline{ Irina~Dymnikova $^{1,2}$, Alexander Sakharov
$^{3,4}$ and J\"urgen~Ulbricht $^{5}$}

\vskip 0.1in

\centerline{\it $^1$ Dep-t of Math. \& Computer Science, Univ. of
Warmia \& Mazury, 10-561 Olsztyn, Poland}

\centerline{\it $^2$ Physico-Technical Institute of the Russian
Academy of Sciences, 194021 St.Petersburg, Russia}

\centerline{\it $^3$ TH Division, PH Department, CERN, CH-1211
Geneva 23, Switzerland}

\centerline{\it $^4$ Department of Physics, Wayne State
University, Detroit, MI 48202, USA}

\centerline{\it $^5$ Swiss Institute of Technology ETH-Z\"urich,
CH-8093 Z\"urich, Switzerland}

 \vskip 0.2in

{\bf Abstract}

\baselineskip=18pt

\noindent Experimental data suggest  the existence of a minimal length scale
in annihilation process for the reaction $\EEGG$. Nonlinear
electrodynamics coupled to gravity and satisfying the weak energy
condition predicts, for an arbitrary gauge invariant lagrangian,
the existence of a spinning charged electromagnetic soliton
asymptotically Kerr-Newman for a distant observer with a
gyromagnetic ratio $g=2$. Its internal structure includes an
equatorial disk of de Sitter vacuum which has properties of a
perfect conductor and ideal diamagnetic, and displays
superconducting behavior within a single spinning soliton.  De
Sitter vacuum supplies a particle with the finite positive
electromagnetic  mass related to breaking of space-time symmetry.
We apply this approach to interpret the existence of a minimal
characteristic length scale in annihilation.

 \vspace*{0.2cm}

\section{Introduction}

The question of intrinsic structure of a fundamental charged
spinning particle such as an electron, has been discussing in the
literature since its discovery by Thomson in 1897. One can roughly
distinguish two approaches. First one deals with point-like
models. In quantum field theory a particle is assumed point-like,
and classical models of the first type consider point-like
particles described by various generalizations of the classical
Hamilton lagrangian $(-mc\sqrt{\dot x\dot x})$ involving higher
derivatives terms or inner variables \cite{att}, and making use of
geometry \cite{geom} or symmetry \cite{symm} constraints. An
elegant recent example is the Staruszkiewicz relativistic rotator
as a fundamental dynamical system whose Casimir invariants are
parameters, but not constants of motion \cite{star}. This gives
rise to a classical model for a point-like relativistic spinning
particle which can be extended to the case when it interacts with
an external electromagnetic field \cite{kass}.

Another type of point-like models of spinning particles goes back
to the Schr\"odinger suggestion that the electron spin can be
related to its Zittebewegung motion \cite{schrod}. The concept of
Zitterbewegung - trembling motion due to the rapid oscillation of
a spinning particle around its classical worldline, has been
worked out in a lot of papers \cite{zbw,pavsic,sm} motivated by
attempts to understand the intrinsic structure of the electron
\cite{bz}. For example, in models based on the Clifford algebras,
the electron is associated with the mean motion of its point-like
constituent whose trajectory is a cylindrical helix (\cite{pavsic}
and references therein).

\vskip0.1in

Second type approach deals with {\it extended} particle models.

The concept of an extended electron, proposed by Abraham
\cite{abr} and Lorentz \cite{lor}, that makes finite the total
field energy, assumed the electron to be a spherical rigid object.
While point-like models typically suffer from an infinite
self-energy, the main problem encountered by extended  models, was
to prevent an electron from flying apart under the Coulomb
repulsion. Theories based on geometrical assumptions about the
"shape" or distribution of a charge density, were compelled to
introduce {\it cohesive} forces of non-electromagnetic origin (the
Poincar\'e stress) testifying that replacing a point charge with
an extended one is impossible within electrodynamics since it
demands introducing cohesive non-electromagnetic forces.

It was clearly formulated by Dirac who  proposed in 1962 the model
of an electron as a charged conducting surface; outside the
surface, the Maxwell equations hold; inside there is no field; a
non-Maxwellian force was assumed as kind of a surface tension, so
the electron is pictured as a spherical bubble in the
electromagnetic field \cite{dirac}.

Similar picture was obtained in the frame of the Dirac non-linear
electrodynamics in the Minkowski space, based on imposing a
nonlinear gauge on a vector potential \cite{dirac51}. The field
equations of this theory have soliton-like solutions which can be
regarded as describing a charged particle \cite{rv}, and admit
further generalization \cite{rodrigues} to yield a classical model
for a spherical charged spinning particle looking as a hole in an
electromagnetic field and demonstrating a solitonic behavior: the
interior of a particle is accessible to any other particle (apart
from electromagnetic repulsion) \cite{rodrigues}.

The Kerr-Newman geometry discovered in linear electrodynamics
coupled to gravity \cite{newman}
$$
ds^2=-dt^2 + \frac{\Sigma}{\Delta}dr^2+ \Sigma d\theta^2
+\frac{(2mr - e^2)}{\Sigma}(dt - a \sin^2\theta d\phi)^2
$$
$$
 + (r^2 + a^2) \sin^2\theta d\phi^2; ~~~~
 A_i = - \frac{e r}{\Sigma}[1; 0, 0, -a \sin^2\theta]
                                                           \eqno(1)
$$
 where $A_i$ is associated electromagnetic potential, and
$$
\Sigma = r^2 + a^2 \cos^2\theta; ~~~ \Delta = r^2 - 2mr + a^2 +
e^2,
                                                           \eqno(2)
$$
have inspired further search for an electromagnetic image of the
electron since Carter \cite{carter} found that the parameter $a$
couples with the mass $m$ to give the angular momentum $J=ma$, and
with the charge $e$ to give an asymptotic magnetic momentum
$\mu=ea$, so that there is no freedom in variation of the
gyromagnetic ratio $e/m$ which is exactly the same as predicted by
the Dirac equation, $g=2$, and it is possible to choose the
parameters in such a way that they agree with the electron
parameters; in the units $\hbar=c=G=1$ we have $~ a=1/2m$, and the
length scale determined by $a$ is about the Compton wavelength
\cite{carter}.

This result suggested  that the spinning electron might be
classically visualized as a massive charged source of the
Kerr-Newman field \cite{werner,lopez}.

The point is that the Kerr-Newman geometry itself cannot model a
particle for the very serious reason discovered by Carter
\cite{carter}: In the case $a^2+e^2> m^2$ appropriate for
modelling a particle since there are no Killing horizons and the
manifold is geodesically complete, just in this case the whole
space is a single vicious set, i.e. such a set in which any point
can be connected to any other point by both a future and a past
directed timelike curve, which means complete and unavoidable
breakdown of causality \cite{carter}.

The Kerr-Newman solution belongs to the Kerr family of the
source-free Maxwell-Einstein equations, the only contribution to a
stress-energy tensor comes from a source-free electromagnetic
field \cite{carter}. It can represent the exterior fields of
spinning charged bodies. The question of an interior material
source for these exterior fields, is the most intriguing question
addressed in a lot of papers. The source models for the
Kerr-Newman interior can be roughly divided into
disk-like\cite{werner,bur74,lopez1},
shell-like\cite{delacruz,cohen,lopez},
bag-like\cite{boyer,trumper,tiomno,bur89,bur2000,behm}, and
string-like (\cite{bur2005} and references therein).
Characteristic radius of a disk is the Compton wavelength
$\lambda_e\simeq{3.9\times 10^{-11}cm}$, and in bag-like models
thickness of an ellipsoid is of order of the electron classical
radius, $r_e\simeq{2.8\times 10^{-13}cm}$.

The problem of matching the Kerr-Newman exterior to a rotating
material source does not have a unique solution, since one is free
to choose arbitrarily the boundary between the exterior and the
interior \cite{werner}.

On the other hand, in nonlinear electrodynamics coupled to gravity
(NED-GR), the field equations admit regular solutions
asymptotically Kerr-Newman for a distant observer, which describe
a spinning electromagnetic soliton (i.e., a regular finite-energy
solution of the nonlinear field equations, localized in the
confined region and holding itself together by its own
self-interaction) \cite{me2006}. Its generic features valid for an
arbitrary nonlinear lagrangian ${\cal L}(F)$ can be outlined
briefly as follows. In NED-GR solutions satisfying the weak energy
condition (non-negative density as measured along any time-like
curve), a spherically symmetric electrically charged soliton has
obligatory de Sitter center in which the electric field vanishes
while the energy density of electromagnetic vacuum achieves its
maximal finite value representing self-interaction \cite{me2004}.
De Sitter vacuum supplies a particle with the finite positive
electromagnetic mass related to breaking of space-time symmetry
from the de Sitter group in the origin \cite{me2004,me2008}. By
the G\"urses-G\"ursey algorithm based on the Newman-Trautman
technique \cite{gurses} it transforms into a spinning
electromagnetic soliton with the Kerr-Newman behavior for a
distant observer. Its internal structure includes the equatorial
disk of a rotating  de Sitter vacuum which has properties of a
perfect conductor and ideal diamagnetic, and displays
superconducting behavior within a single spinning particle
\cite{me2006}.

Experimental limits on size of a lepton \cite{size} are much less
than its Compton wavelength and classical radius. This suggests
that an extended fundamental particle can have one more,
relatively small characteristic length scale, related to gravity.

To get an evidence for an extended particle picture, we worked out
data of experiments performed to search for compositeness or to
investigate a non-point-like behavior, with focus on
characteristic energy scale related to characteristic length scale
of interaction region \cite{size4}.

 In this paper we outline the
experimental results on the QED reaction measuring the
differential cross sections for the process $ \EEGG $ at energies
from $\sqrt{s} $=55~GeV to 207 GeV using the data collected with
the VENUS, TOPAZ, ALEPH, DELPHI L3 and OPAL from 1989 to 2003.
Experimental data suggest  the existence of a minimal length scale
in annihilation reaction $\EEGG$. The global fit to the data is 5
standard deviations from the standard model expectation for the
hypotheses of an excited electron and of contact interaction with
non-standard coupling \cite{size3}, corresponding to the cut-off
scale $E_{\Lambda} =1.253$~TeV and to related characteristic
length scale $l_e\simeq{1.57 \times 10^{-17}}$~cm. We interpret
this experimental effect by applying theoretical results obtained
in nonlinear electrodynamics coupled to gravity.

\section{Experimental evidence for an extended lepton}

The purely electromagnetic interaction $ \EEGG $ is ideal to test
 QED because it is not interfered by the $ Z^{o} $ decay. This
reaction proceeds via the exchange of a virtual electron in the t
- and u - channels, while the s - channel is forbidden due to
angular momentum conservation. Differential cross sections for the
process $ \EEGG $, are measured at energies from  $\sqrt{s}
$=55~GeV to 207 GeV using the data collected with the VENUS
\cite{VENUSdiffCrossSection}, TOPAZ \cite{TOPAZdiffCrossSection},
ALEPH \cite{ALEPHdiffCrossSection}, DELPHI
\cite{DELPHIdiffCrossSection}, L3 \cite{L3diffCrossSection} and
OPAL \cite{OPALdiffCrossSection} detector from 1989 to 2003.

Comparison of the data with the QED predictions are used to
constrain models with an excited electron of mass $ m_{e^{*}} $
replacing the virtual electron in the QED process
\cite{ExcitedElectron}, and a model with deviation from QED
arising from an effective interaction with non-standard $ e^{+}
e^{-} \gamma $ couplings and $ e^{+} e^{-} \gamma \gamma $ contact
terms \cite{DirectContact}.

A heavy excited electron could couple to an electron and a photon
via magnetic interaction with an effective lagrangian \cite{LITKE}

$$
{\mathcal L}_{{excited}}=\frac{e\lambda}{2m_{e^{*}}}
\overline{\psi_{e^{*}}}\sigma_{\mu\nu}\psi_{e}F^{\mu\nu}
                                                                 \eqno(3)
$$
Here $ \lambda $ is the coupling constant, $ F^{\mu\nu} $ the
electromagnetic field, $ \psi_{e^{*}} $ and $ \psi_{e} $ are the
wave function of the heavy electron and the electron respectively;
$ \lambda $ and $ m_{e^{*}} $ are the model parameters.
Differential cross-section involves a deviation term $
\delta_{new} $ from the QED differential cross-section including
radiative effects up to $ O(\alpha^{3}) $. The modified equation
reads
$$
(d\sigma/d\Omega)_{theo}=(d\sigma/d\Omega)_{O(\alpha^{3})}
(1+\delta_{new})
                                                              \eqno(4)
$$

If the center-of-mass energy $\sqrt{s}$ satisfies the condition $
s/m^{2}_{e^{*}} << 1$, then $\delta_{new}$ can be approximated as
$$
\delta_{new}= s^{2}/2 ( 1/\Lambda^{4} ) (1-\cos^{2}\Theta)
                                                                    \eqno(5)
$$

In this approximation, the parameter $\Lambda $ is the QED cut-off
parameter, $ \Lambda^{2}=m^{2}_{e^{*}}/\lambda $. In the case of
arbitrary $\sqrt{s}$ the full equation of ref.\cite{LITKE} is used
to calculate $ \delta_{new} = f(m_{e^{*}}) $. The angle $ \Theta $
is the open angle of the two most energetic photons emitted  with
angles $ \Theta_{1} $ and $ \Theta_{2} $ with respect to the beam
axis defined below
$$
 \mid cos ( \Theta ) \mid = 1/2 ( \mid cos ( \Theta_{1} ) \mid +
                     \mid cos (2\pi - \Theta_{2})\mid)
                                                                   \eqno(6)
$$

The third order QED differential cross section is calculated
numerically  up to $ O(\alpha^{3}) $, by generating a high number
of Monte Carlo  $ \EEGG $ events
\cite{GAMMAgenerator,L3diffCrossSectionB}. The angular
distribution of these events was fitted with a high order
polynomial function to get an analytical equation for the cross
section as function of the scattering angle defined in (6).

\begin{figure}
\vspace{-5.0mm}
\begin{center}
 \epsfig{file=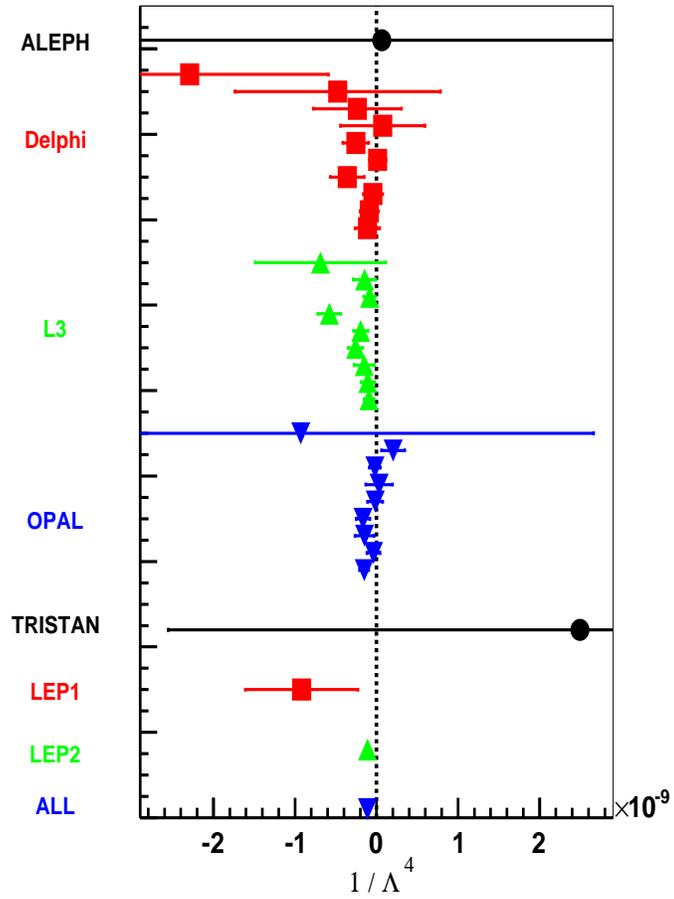,width=9.0cm,height=12.7cm}
\end{center}
\caption{The $\chi^{2}$ minima for all $ 1/\Lambda^{4} [{\rm
GeV}^{-4}] $ values. } \label{plotfitRESULTS}
\end{figure}

An overall $ \chi^{2} $ test between 55 GeV and 207 GeV was
performed on the published differential cross sections. The single
results of the different $ 1/\Lambda^{4} [ 1/ {\rm GeV}^{4} ] $
minima are displayed  in Fig.{\ref{plotfitRESULTS}}. The upper
part shows the 4 LEP experiments and the lower part shows the
 combined in three groups results from TRISTAN, LEP 1, LEP 2, and
 the overall result of $1/\Lambda^4=-(1.11\pm 0.70)\times{10^{-10}
 {\rm GeV}^{-4}}$.

Systematic errors arise from the luminosity evaluation, from the
selection efficiency, background evaluations, the choice to use
the Born level or $ \alpha ^{3} $ theoretical QED cross section as
reference cross section, the choice of the fit procedure, the
choice of the fit parameter and the choice of the scattering angle
$ | cos \Theta | $ in particular in comparison between data and
theoretical calculation.

The maximum estimated error for the value of the fit from the
luminosity, selection efficiency and background evaluations is
approximately $ \delta \Lambda/\Lambda = 0.01 $
\cite{SystematicERROR}. The choice of the theoretical QED cross
section was studied with 1882 [$\EEGG$] events from the L3
detector \cite{SystematicERROR}. In Fig.
{\ref{plot_QEDcrossBornALPHA3}} the measured data points of the $
\EEGG $ reaction are shown together with the QED Born and the $
\alpha ^{3} $ level approximations. In part b) the sensitivity of
the measured data points to QED cross sections is visible.

A drop in the $ \chi^{2} $ by approximately a factor two favors
the QED $ \alpha ^{3} $ level to be used for the fit. For a small
sample of $ \EEGG $ events the fit values $ \Lambda $ are compared
for $ \chi^{2} $, Maximum-Likelihood, Smirnov-Cramer von Misis,
Kolmogorov test, all with and without binning
\cite{ERRORfitMETHOD}. An approximately $ \delta \Lambda /\Lambda
= 0.005 $ effect is estimated for the overall fit with the fit
parameter $ P = ( 1/\Lambda^{4} )$. The $\chi^2$ overall fit
displays a minimum in the $\chi^{2}$ as we see in
Fig.{\ref{plot-paramterLAMDA}).

The use of different definitions of scattering angles
\cite{TOPAZdiffCrossSection} introduces in the $ \mid cos ( \Theta
) \mid $ an error of approximately $ \delta \mid cos ( \Theta )
\mid  = 0.0005 $. In the worst case of  scattering angles close to
$ 90^{o}, $ the $ | cos ( \Theta ) |_{experiment} \sim 0.05  $
would result in $ ( \delta \Lambda /\Lambda )_{\delta
|\cos(\Theta)|} = 0.01 $. The total systematic error is $ \delta
\Lambda /\Lambda \approx 0.015 $.

\begin{figure}
\vspace{-2.0mm}
\begin{center}
 \epsfig{file=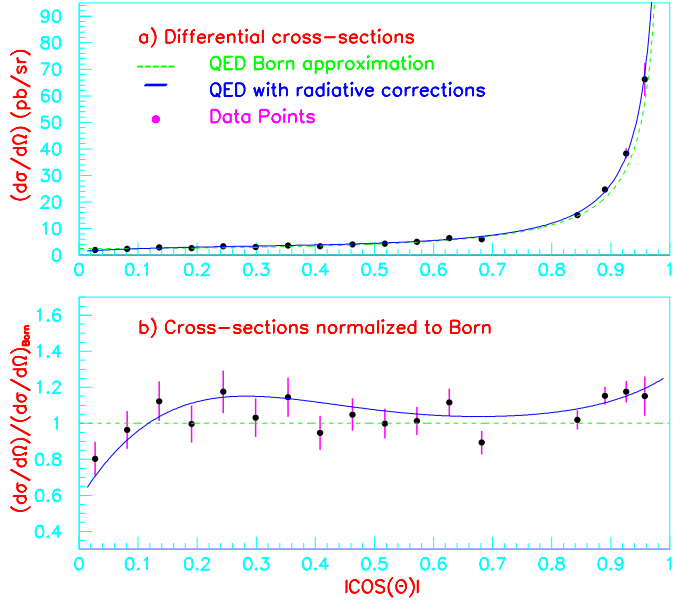,width=10.1cm,height=10.07cm}
\end{center}
\caption{QED cross section and experimental data }
\label{plot_QEDcrossBornALPHA3}
\end{figure}

The hypothesis used in (3) and (4) assumes that an excited
electron will increase the total QED-$ \alpha^{3} $ cross section
and change the angular distribution of the QED cross section.
Contrary to these expectations, the fit expresses a minimum with a
negative fit parameter $ 1/\Lambda^{4} $ of a significance of
approximately $5\times \sigma $ .

For an effective contact interaction with non-standard coupling, a
cut-off parameter $ \Lambda_{C} $ is introduced to describe the
scale of interaction with the lagrangian \cite{DirectContact}

$$
{\mathcal L}_{{contact}}=i\overline{\psi_{e}}
\gamma_{\mu}(D_{\nu}\psi_{e})
\left(\frac{\sqrt{4\pi}}{\Lambda^{2}_{C6}}F^{\mu\nu}+
\frac{\sqrt{4\pi}}{\tilde{\Lambda}^{2}_{C6}}\tilde{F}^{\mu\nu}\right)
                                                                           \eqno(7)
$$

The effective Lagrangian chosen in this case has an operator of
dimension 6, the wave function of the electrons is $ \psi_{e} $,
the QED covariant derivative is $ D_{\nu} $, the tilde on $
\tilde{\Lambda}_{C6} $ and $ \tilde{F}^{\mu\nu} $ stands for
duals. As in the case of excited electron the corresponding
differential cross section involves a deviation term $
\delta_{new} $ from the QED differential cross section including
radiative effects up to $ O(\alpha^{3}) $, and $\delta_{new}$
reads as
$$
\delta_{new}=s^{2}/(2\alpha)(1/\Lambda^{4}_{C6} +
1/\tilde{\Lambda}^{4}_{C6}) (1-\cos^{2}\Theta)
                                                                             \eqno(8)
$$

The angle $ \Theta $ is the angle of the emitted photons with
respect to the beam axis defined in (6). For the fit procedures
discussed below we set
$\Lambda_{C6}=\tilde{\Lambda}_{C6}=\Lambda_{C} $.

\begin{figure}[htbp]
\vspace{-5.0mm}
\begin{center}
\rotatebox{180}{}
 \epsfig{file=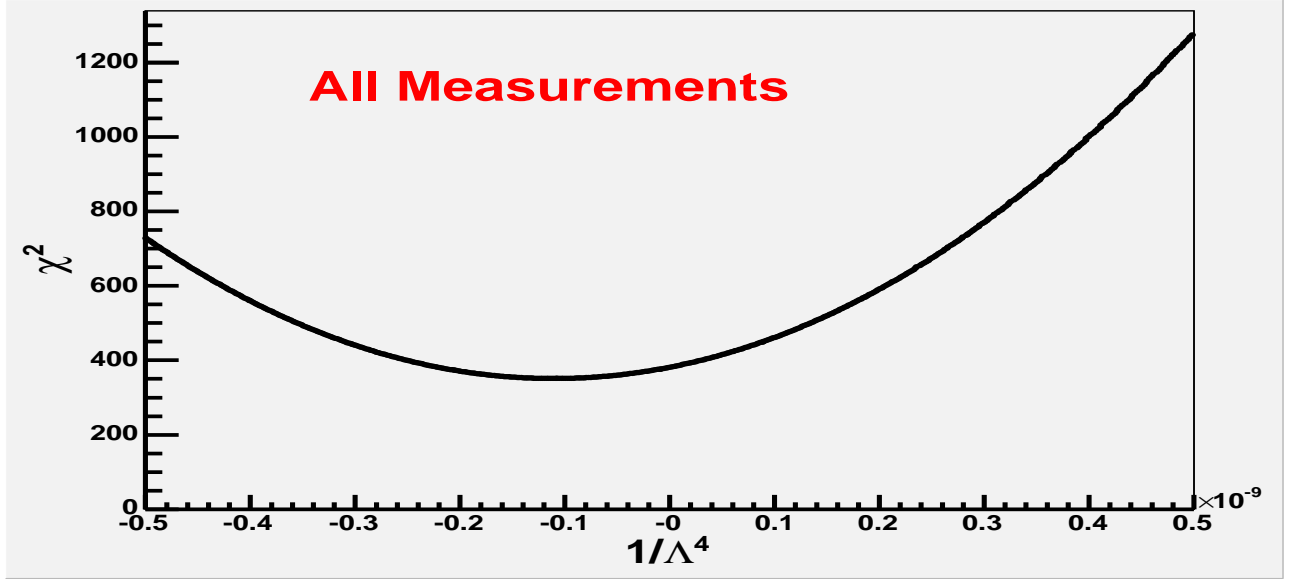,width=17cm,height=7.7cm}
\end{center}
\caption{A minimum in the $\chi^{2}$ for the overal fit ($ 1/\Lambda^{4} [{\rm
GeV}^{-4}] $).}
\label{plot-paramterLAMDA}
\end{figure}

The $ \chi^{2} $ fit for the hypothesis of the excited electron,
eq.(3), was repeated for the hypothesis of the effective contact
interaction, eq.(7), using $ ( 1/\Lambda^{4}_{C} ) $ as fit
parameter. As in the hypothesis of the excited electron also for
the effective contact interaction, an increase of the total
QED-$\alpha^{3} $ cross section and a change of the angular
distribution were expected. In contrary to both hypothesis also
the best fit value of all data $ ( 1/\Lambda^{4}_{C}
)_{best}=-(4.05\pm0.73)\times 10^{-13}{\rm GeV} ^{-4} $ is
negative with significance about $ 5 \times \sigma $. The fit does
not allow to distinguish between both above hypothesis. The
results indicate decreasing cross section of the process $ \EEGG $
with respect to that predicted by pure QED. The calculation of the
QED-$ \alpha^{3} $ cross section assumes a scattering center as a
point. If the electron is an extended object, its structure would
modify the QED cross section if the test distances (CM-scattering
energy) are smaller than its characteristic size.

It is remarkable that for both hypothesis the excited electron and
effective contact interaction, the $ \chi^{2} $ test leads to a
best fit value $ (1/\Lambda^{4})_{best} $ and $
(1/\Lambda^{4}_{C})_{best} $ for the complete data set  with a
$5\sigma$ significance.

 With the best value  $ (1/\Lambda)^{4}_{C} $ one
can calculate the energy scale $E_{\Lambda}= (\Lambda_C)_{best} =
1.253$ TeV \cite{size3} which corresponds to a length scale  $ l_e
\simeq 1.57 \times 10^{-17} $ cm as the distance of the closest
approach of particles which cannot be made smaller and suggests
the existence of a minimal characteristic length scale in
annihilation.

\section{Electromagnetic soliton}

In the nonlinear electrodynamics minimally coupled to gravity
(NED-GR), the action is given by (in geometrical units $ G = c = 1
$)
$$
       S = \frac{1}{16\pi} \int d^{4}x \sqrt{-g}
           ({\cal R} - {\mathcal{L}}(F));~~
           F = F_{\mu\nu} F^{\mu\nu}
                                                                            \eqno(9)
$$
where ${\cal R}$ is the scalar curvature. The gauge-invariant
electromagnetic Lagrangian $ {\mathcal{L}}(F) $ is an arbitrary
function of $ F $ which should have the Maxwell limit, $
{\mathcal{L}} \rightarrow F $,  in the weak field regime.

In the case of electrically charged structure, a field invariant $
F $ must vanish for $ r \rightarrow 0 $ to guarantee regularity
\cite{kirill}, and the electric field strength is zero in the
center of any regular charged NED-GR structure. The field
invariant $ F $ vanishes at both zero and infinity where it
follows the Maxwell weak field limit. In both limits $F\rightarrow
-0$, so that $ F $ must have at least one minimum in between,
where an electrical field strength has an extremum too
\cite{kirill,me2004}.

A stress-energy tensor of a spherically symmetric electromagnetic
field $\kappa T^{\mu}_{\nu}= -2 {\cal L}_F
F_{\nu\alpha}F^{\mu\alpha}+\frac{1}{2}\delta_{\nu}^{\mu} {\cal L}
$, where $\kappa=8\pi G$, has  the algebraic structure
$$
       T^{t}_{t} = T^{r}_{r}
                                                                    \eqno(10)
$$
Symmetry of a source term  leads to the metric \cite{me2000}
$$
ds^2 = g(r) dt^{2} - \frac{dr^{2}}{g(r)} - r^{2} d\Omega^{2}
                                                                        \eqno(11)
$$
The metric function and mass function are given by
$$
       g(r) = 1 - \frac{2{\cal{M}}(r)}{r}:~~~
       {\cal{M}}(r) = \frac{1}{2}\int\limits_{0}^{r} \rho(x)x^{2} dx
                                                                              \eqno(12)
$$

For the class of regular spherical symmetric geometries with the
symmetry of a source term given by (10), the weak energy condition
 leads inevitably to de Sitter asymptotic at approaching a
regular center \cite{me2000,me2002}
$$
     p=-\rho;~~ ~~   g(r) = 1 - \frac{\Lambda}{3} r^{2}
                                                                                  \eqno(13)
$$
with cosmological constant $ \Lambda = 8 \pi \rho_0$ where
$\rho_0=\rho(r=0)$ is the finite density in the regular center. As
a result, the mass of an object described by (10)-(12), $m={\cal
M}(r\rightarrow\infty)$, is generically related to breaking of
space-time symmetry from the de Sitter group in the origin, and to
de Sitter vacuum trapped inside \cite{me2002}.

Regular electrically charged spherically symmetric solutions
describe an electromagnetic soliton with the obligatory de Sitter
center in which field tension goes to zero, while the energy
density of the electromagnetic vacuum $T_t^t$ achieves its maximal
finite value which represents the de Sitter cutoff for the
self-interaction divergent for a point charge \cite{me2004}.

For a distant observer, it is described by the
Reissner-Nordstr\"om asymptotic
$$
g(r)=1-\frac{r_g}{r} + \frac{e^2}{r^2}
                                                                             \eqno(14)
$$
where $r_g=2m$ is the Schwarzschild gravitational radius.

 For all solutions specified by (10), there exists the
surface of zero gravity at which the strong energy condition
($\rho + \sum p_{k} \ge 0$) is violated  which means that
gravitational acceleration changes its sign and becomes repulsive
\cite{me96,me2000}.

\vskip0.1in

Spherically symmetric solutions satisfying the condition (10)
belong to the Kerr-Schild class \cite{behm,ssqv}. By the
G\"urses-G\"ursey algorithm \cite{gurses} they can be transformed
into regular solutions describing a spinning charged soliton. In
the Boyer-Lindquist coordinates the metric is
$$
ds^2 = \frac{2f - \Sigma}{\Sigma} dt^2  + \frac{\Sigma}{\Delta}
dr^2 + \Sigma d\theta^2 - \frac{4af\sin^2\theta}{\Sigma}dt d\phi +
\biggl(r^2 + a^2 +
\frac{2fa^2\sin^2\theta}{\Sigma}\biggr)\sin^2\theta d\phi^2
                                                            \eqno(15)
$$
$$
\Sigma=r^2+a^2\cos^2\theta;~~\Delta = r^2 + a^2 - 2f(r)
                                                                     \eqno(16)
$$

The function $f(r)$ in (15) is given by
$$
f(r)=r{\cal M}(r)
                                                                   \eqno(17)
$$
where density profile in (12) is that for a nonlinear spherically
symmetric electromagnetic field.

For NED-GR solutions satisfying the weak energy condition, ${\cal
M}(r)$ is everywhere positive function growing monotonically  from
${\cal M}(r)=4\pi\rho_0r^3/3$ as $r\rightarrow 0$ to $m$ as
$r\rightarrow\infty$. The mass  $m$, appearing in a spinning
solution, is the finite positive electromagnetic mass
\cite{me2006,me2004}.

The condition of the causality violation \cite{carter} takes the
form \cite{me2006}
$$
r^2 + a^2 +\Sigma^{-1}2f(r)a^2\sin^2\theta < 0
                                                                    \eqno(18)
$$
and is never satisfied due to non-negativity of the function
$f(r)$.

In the geometry with the line element (15), the surfaces $r =
const$ are the oblate ellipsoids
$$
r^4-(x^2+y^2+z^2-a^2)r^2-a^2 z^2=0
                                                                 \eqno(19)
$$
which degenerate, for $r = 0$, to the equatorial disk
$$
x^2 + y^2 \leq a^2, ~~ z = 0
                                                                 \eqno(20)
$$
centered on the symmetry axis.

For a distant observer, a spinning electromagnetic soliton is
asymptotically Kerr-Newman, with $f(r)=mr-e^2/2$, and the
gyromagnetic ratio $g=2$.

For $r\rightarrow 0$ the function $f(r)$ in (15) approaches  de
Sitter asymptotic
$$
2f(r)=\frac{r^4}{r_0^2}; ~~~ ~~ r_0^2=\frac{3}{\kappa\rho_0}
                                                               \eqno(21)
$$
and the metric  describes rotating de Sitter vacuum in the
co-rotating frame \cite{me2006}.

In the NED-GR regular solutions, an internal equatorial disk (20)
is filled with rotating de Sitter vacuum,  it has properties of
both perfect conductor and ideal diamagnetic, and displays
superconducting behavior within a single spinning soliton
\cite{me2006}.

\section{Origin of a minimal length scale in annihilation}

The minimum in the fit found with $5\times\sigma$ significance,
corresponds to the characteristic length scale
$l_e\simeq{1.57\times10^{-17}}$ cm related to the energy scale
$E_{\Lambda}\simeq{1.253}$ TeV.

The existence of the limiting length scale $l_e$ in experiments on
annihilation, testifies for an extended particle rather than a
point-like one. The effective size of an interaction region $l_e$
corresponds to a minimum in $\chi^2$, so that it can be understood
as a minimal length scale in annihilation which cannot
 be made smaller.

 Generic features of electromagnetic soliton  give some idea
 about the origin of the characteristic length scale $l_e$
 given by experiments.  The certain feature of annihilation process is that
at a certain stage a region of interaction is neutral and
spinless. We can roughly model it by a spherical lump with de
Sitter vacuum interior. The key point is the existence of
zero-gravity surface at which strong energy condition is violated
\cite{me96,me2000} and gravitational acceleration becomes
repulsive. The related length scale $r_*\simeq{(r_0^2r_g)^{1/3}}$
appears naturally in direct matching de Sitter interior to the
Schwarzschild exterior \cite{werner1}.

The gravitational radius of a lump on the characteristic energy
scale $E_{\Lambda}\simeq{1.25}$ TeV, is $r_g\simeq{3.32\times
10^{-49}}$ cm. Adopting for the interior de Sitter vacuum the
experimental vacuum expectation value for the electroweak scale
$E_{EW}=246$
 GeV related to the electron mass \cite{okun} we get the de Sitter
horizon radius $r_0={1.374}$ cm. Characteristic radius of zero
gravity surface is $r_*\simeq{{0.86\times 10^{-16}}}$ cm, so that
the scale $l_e$ fits inside a region where gravity is already
repulsive. The scale $l_e$ can be imagined as a distance at which
electromagnetic attraction is stopped by gravitational repulsion
due to interior de Sitter vacuum.

In extended regular models based on nonlinear electrodynamics
there exists a characteristic cutoff on self-energy whose value
depends on a chosen density profile (\cite{me2004} and references
therein). In regular models with de Sitter interior it can be
qualitatively evaluated as
$$
\frac{e^2}{r_e^4}\simeq 8\pi G\rho_0=\frac{3}{r_0^2}
                                                      \eqno(22)
$$
It gives a rough estimate for the characteristic length scale
$r_e$ at which electromagnetic attraction is balanced by de Sitter
gravitational repulsion $ r_e={1.05\times 10^{-17}}cm$ which is
quite close to experimental value $l_e$, although estimate is
qualitative and model-independent.

\section{Summary}

Nonlinear electrodynamics coupled to gravity predicts that
spinning particles dominated by the electromagnetic interaction,
would have to have de Sitter interiors arising naturally in the
regular geometry asymptotically Kerr-Newman for a distant
observer. De Sitter vacuum supplies a particle with the finite
positive electromagnetic mass related to breaking of space-time
symmetry \cite{me2006}.

 In all asymptotically Kerr-Newman models,
 symmetry of an oblate ellipsoid (3) leads to estimates of the intrinsic radius of an
internal disk by the Compton wavelength $\simeq {3.9\times
10^{-11}} ~cm$, and of the transverse size (thickness of
ellipsoid) by the classical electron radius $\simeq {2.8\times
10^{-13}} ~cm$.

NED theories appear as low-energy effective limits in certain
models of string/M-theories (for review \cite{fradkin,witten}).
The above results apply to the cases when the relevant
electromagnetic scale  is much less than the Planck scale.

 Experiments reveal the existence of a minimal length scale
in the process of annihilation,  $l_e\simeq{1.57 \times
10^{-17}}~cm$ for the electron. This characteristic length can be
explained as a distance at which electromagnetic attraction in
annihilation is stopped by gravitational repulsion due to an
interior de Sitter vacuum.

One can conclude that experiments suggest an extended electron
picture.

\section*{Acknowledgement}

We are very grateful to Andr\'e Rubbia for encouragement and
helpful remarks. This work was supported by University of Warmia
and Mazury through sponsorship for I.D. at CERN.

\end{document}